# Superconducting Dome in $La_{3-x}Sr_xNi_2O_{7-\delta}$ Thin Films

Maosen Wang,[1,2,*] Bo Hao,[1,2,*] Wenjie Sun,[1,2,*] Shengjun Yan,[1,2] Shengwang Sun,[1,2] Hongyi Zhang,[1,2] Zhengbin Gu,[1,2] and Yuefeng Nie[1,2,3,†]

[1]National Laboratory of Solid State Microstructures, Jiangsu Key Laboratory of Artificial Functional Materials, College of Engineering and Applied Sciences, Nanjing University, Nanjing 210093, China.

[2]Collaborative Innovation Center of Advanced Microstructures, Nanjing University, Nanjing 210093, China.

[3]Jiangsu Physical Science Research Center, Nanjing 210093, China.

## Abstract

The ambient-pressure superconductivity in $La_3Ni_2O_7$ thin films via compressive epitaxial strain provides a highly accessible platform for diverse characterization techniques, facilitating the studies of high-temperature superconductivity. Here, we systematically map the phase diagram and reveal the superconducting dome with an electron-hole crossover in compressively strained $La_{3-x}Sr_xNi_2O_{7-\delta}$ thin films by simultaneously tuning Sr doping and oxygen content. The maximum transition temperature ($T_c$) coincides with an anomalous sign change in the Hall coefficient ($R_H$), reminiscent of electron-doped cuprates, which may signal a Fermi surface reconstruction. Beyond the superconducting dome, a $\ln 1/T$ insulating regime and a $T$-linear resistivity regime are also resolved, resembling behaviors observed in cuprates and infinite-layer nickelates. This work reveals a dome-shaped relationship between $T_c$ and $R_H$ and establishes a key framework for understanding unconventional superconductivity in nickelate systems.

## Main

A defining feature of unconventional superconductors lies in their complex phase diagram where a doping-dependent superconducting dome emerges in proximity to multiple competing orders encompassing pseudogap, antiferromagnetic order, strange metal, and density wave orders [1-4]. Despite decades of extensive investigation into the high-superconducting transition temperature ($T_c$) cuprates as model systems, the fluctuations of the competing orders and their interplay with the superconducting phase remain elusive [5-8]. The recently discovered high-$T_c$ superconductivity in Ruddlesden-Popper (RP) nickelates ($R_{n+1}Ni_nO_{3n+1}$, $R$ = rare-earth) under high pressure, coupled with its stabilization at ambient pressure through epitaxial compressive strain in thin films, has established a novel testbed for elucidating these open questions [9-22]. Among the RP nickelates, the bilayer $La_3Ni_2O_7$ hosts a nominal transition metal $3d^{7.5}$ electronic configuration [9, 17], which is distinct from the $3d^9$ configuration in both undoped cuprates and infinite-layer nickelates [23, 24]. Furthermore, the contributions of both Ni $3d_{x^2-y^2}$ and $3d_{z^2}$ orbitals to the low-energy physics lead to a multiband electronic structure, with the strong interlayer $d_{z^2}$ orbital coupling via the O-2*p* orbitals of the inner apical oxygen proposed to be crucial for superconductivity [9, 25-38]. Notably, the absence of long-range magnetic order in bilayer $La_3Ni_2O_7$ also contrasts sharply with cuprates [17, 39-41], where superconductivity arises from the antiferromagnetic ground state through carrier doping [2, 5, 6]. These distinctions raise the issue of whether analogous interplay between superconducting phase and potential competing orders exists in the doping-dependent phase diagram of bilayer nickelates as in cuprates and infinite-layer nickelates. However, a systematic doping-dependent investigation is largely absent to date on account of the great challenges in synthesizing the superconducting bilayer nickelates.

Heterovalent cation doping has been demonstrated to be a powerful method to map the evolution of superconductivity and electronic states with carrier density in unconventional superconductors, paradigms including hole-doped cuprates and infinite-layer nickelates via alkaline-earth metals [42-45] and electron-doped cuprates through $Ce^{4+}$ [46]. Anion regulation, on the other hand, provides another essential approach to tune the carrier density especially in oxide superconductors, most notably exemplified by the modulation of superconductivity through oxygen concentration in cuprates [47-51]. In bilayer nickelates, the oxygen concentration can be readily tuned [14, 19-22], offering an effective knob to investigate the impact of carrier density on superconductivity. Meanwhile, no superconductivity has been observed yet in heterovalent cation doping efforts for bilayer nickelates in either bulk crystals or thin films [52-54], underscoring its challenges.

In this letter, we report the dependence of superconductivity on Sr doping and oxygen content in compressively strained $La_{3-x}Sr_xNi_2O_{7-\delta}$ thin films. Given the multiband structure, we employ the Hall coefficient ($R_H$) to track carrier density evolution, revealing a universal superconducting dome in $T_c$ vs $R_H$ phase space, accompanied by $T$-linear resistivity and $\ln 1/T$ insulating behavior in the normal state. An anomalous electron-hole crossover marked by the sign change of $R_H$ is observed near the maximum of $T_c$, which is reminiscent of the electron-doped cuprates.

A series of $La_{3-x}Sr_xNi_2O_{7-\delta}$ thin films with $SrTiO_3$ capping layer serving as structural support were grown on single-crystalline (001) $SrLaAlO_4$ substrates via reactive molecular-beam epitaxy (MBE) with various Sr doping levels. The quality of the films was confirmed by x-ray diffraction (XRD) $2\theta$-$\omega$ scans along the (00$l$) direction (Fig. S1) [55]. To tune the oxygen content, we employ strong annealing conditions to achieve excessively oxygen doped state, followed by systematically reducing the oxygen content through repeated *in situ* vacuum annealing cycles. An optimized post-growth ozone annealing pathway was adopted to suppress second-phase formation, as suggested in Ref. [22]. Additional experimental details are provided in Supplementary Materials [55].

We first introduce the general electrical transport properties of a representative sample with $x$ = 0.21 (Fig. 1). Owing to the extreme sensitivity to oxygen content in this compound, the transport properties of the as grown films can vary to some extent, as shown in Fig. S8 [55]. After an initial excess ozone annealing to achieve the excessively doped state as mentioned above, the film exhibits superconducting transition with relatively low resistivity in normal state, while subsequent *in situ* vacuum annealing leads to higher resistivity. As the film tends to be oxygen-deficient, a resistivity minimum accompanied by a weakly insulating behavior becomes more obvious, while the superconducting transition persists. The initial enhancement of $T_{c,98\%}$ during early vacuum annealing cycles marked in Fig. 1(b) followed by further suppression indicates an oxygen-controlled crossover from higher doped state to lower doped state, suggesting a dome-like relationship between $T_c$ and the dopant concentration tuned by the oxygen content. Here and for the rest of the paper, $T_{c,98\%}$ is defined as the temperature at which the resistivity drops to 98% of the extrapolated normal state resistivity, linearly fitted near the superconducting transition. The similar peak of $T_c$ is also confirmed in other samples, as shown in Fig. S3 [55]. When the oxygen is further depleted, the superconducting transition becomes insufficient and a superconductor-insulator transition occurs.

To evaluate the effect of oxygen content variations, the Hall coefficient $R_H$ is considered [Fig. 2(a)]. $R_H$ was measured down to 50 K, a temperature close to $T_c$ at which the Hall measurements remain unaffected by the onset of superconducting fluctuations. With the resistivity $\rho$ at 200 K increasing monotonically as oxygen depletion progresses, $R_H$ at 50 K also follows a monotonic increase with $\rho$ at 200 K [Fig. 2(b)], accompanied with a sign change from negative to positive. At 200 K, the $R_H$ remains negative across all oxygen content, with its magnitude $|R_H|$ exhibiting a relatively gentle evolution upon oxygen reduction. The $|R_H|$ is comparable with prior reports in $La_3Ni_2O_7$ and $La_{2.85}Pr_{0.15}Ni_2O_7$ films [20, 21], but distinctly lower than those in cuprates and infinite-layer nickelates [44, 45, 57]. Considering the simplest approximation of $R_H = 1/ne$, where $n$ is the carrier density and $e$ is the electron charge, the evolution of $|R_H|$ at 200 K contradicts the single-band model, in which oxygen incorporation results in hole doping and predicts an increasing $|R_H|$. Combined with the zero-crossing to negative of $R_H$ at 50 K in the oxygen-rich region (discussed later), these anomalous $R_H$ behaviors indicate a multiband electronic structure at the Fermi level, consistent with recent angle-resolved photoemission spectroscopy (ARPES) results in thin films and bulk crystals [32, 37, 38]. Similar $R_H$ anomalies are also observed in other samples, as shown in Fig. S4 [55].

Despite the complexity of deducing the precise carrier density from $R_H$ owing to the multiband nature of the Fermi surface, the monotonic dependence between $R_H$ (50 K) and oxygen doping level enables tracking carrier density evolution and establishing a $T$-$R_H$ phase diagram. First, the normal state resistivity behavior was carefully examined, with one representative $\rho(T)$ curve depicted in Fig. 1(c). A notable feature for the resistivity at high temperature is the $\rho \propto T$ linearity, which remains a long-standing puzzle in high-$T_c$ cuprates and many other strongly correlated systems [2, 4, 58-62].

When it comes to the lower temperature, $\rho$ deviates from the $T$-linear scaling and undergoes a metal-insulator transition (MIT). The resistivity upturn followed above $T_c$ exhibits a logarithmic divergence with temperature, resembling observations in other unconventional superconductors such as underdoped cuprates and infinite-layer nickelates [4, 63-66].

By tracking characteristic temperature scales across vacuum annealing cycles, the evolution of resistivity in $La_{2.79}Sr_{0.21}Ni_2O_7$ thin films is resolved in Fig. 3. The normal state $T$-$R_H$ phase diagram is dissected into a metallic region and a weakly insulating region by the metal-insulator transition temperature $T_{\mathrm{MIT}}$, with a superconducting dome emerging below $T_{c,98\%}$. As $R_H$ (50 K) increases, the weakly insulating region gradually extends to higher temperatures, eventually dominating the normal state up to 200 K. Notably, two regimes of behavior are resolved: a broad range of $\rho \propto T$ at high temperature in the metallic state and $\rho \propto \ln 1/T$ above the superconducting dome in the weakly insulating state. The temperature $T_1$, defined as the characteristic temperature at which $\rho$ deviates from the $T$-linear behavior, is relatively insensitive to $R_H$ (50 K) with $T_1$ ~120 K, which markedly differs from the asymmetric fan-shaped regime of $T$-linear resistivity observed in cuprates and infinite-layer nickelates [7, 60, 67]. On the other hand, the upper boundary of the $\ln 1/T$ insulating behavior, denoted by the characteristic temperature $T_2$, exhibits a dome-like evolution with $R_H$ (50 K).

We now turn to the superconducting dome in $La_{3-x}Sr_xNi_2O_{7-\delta}$ thin films. In previous studies, we have proposed a Sr doping dependent phase diagram [68], while in this work, after systematically varying the Sr doping level ($0 \le x \le 0.21$) and oxygen content, a more general superconducting dome ($T_c$ vs $R_H$) is mapped in Fig. 4. Albeit through different mechanisms—chemical substitution versus oxygen content control, the resembling tuning effects to carrier density between Sr substitution and oxygen incorporation make $R_H$ a unifying parameter to track the evolution of $T_c$ across different doping routes and uncover the intrinsic properties in bilayer nickelate films. Despite variations in Sr doping level and uncertainties in oxygen content arising from growth and the post annealing processes, $T_{c,98\%}$ of the respective sample follows an analogous evolution with $R_H$ (50 K), revealing a universal superconducting dome. Note that the onset $T_c$ is relatively insensitive to some imperfections on the superconducting transition, offering a robust foundation for determining the dome. Interestingly, a small dip of $T_c$ was observed in some samples, which is similar to the local minimum of $T_c$ in $La_{2-x}Sr_xCuO_4$ and $La_{2-x}Ba_xCuO_4$ attributed to the stripe order near $x = 1/8$ [42, 43]. The similar dip has also been reported in infinite-layer nickelates, with its significance remains unclear [44, 45]. However, the reproducibility of the dip was limited across samples, calling for further investigation to clarify its underlying mechanism.

To investigate the anomalous behavior of $R_H$ and the multiband nature mentioned before, we adopt a simplified two-band model with both electron-like pocket and hole-like pocket at the Fermi surface as previously applied to the infinite-layer $Nd_{1-x}Sr_xNiO_2$ superconductors [44, 45]. The temperature $T_H$, at which the $R_H$ changes sign for the samples with $x = 0.21$, has separated the phase diagram into electron- and hole-dominated regions, as illustrated in Fig. 4(a). Taking the dual carrier contributions into account, the sign changes in the $R_H$ depicted in Fig. 2 and S4 can be attributed to the divergent temperature-dependent hole and electron mobility, combined with the distinct doping responses of hole and electron density [44, 45]. This two-band model is further corroborated by electronic structure calculations and ARPES results in both bulk crystals at ambient pressure and thin films of bilayer nickelates, with an electron-like $\alpha$ sheet and a hole-like $\beta$ sheet deriving from the Ni-3$d_{x^2-y^2}$ orbital on the Fermi surface [9, 25-38]. It should be noted that the $R_{xy}$ in our measurements exhibits a completely linear dependence on the magnetic field up to 9 T (Fig. S5) [55], preventing the quantitative estimation of the carrier densities and mobilities through the two-band model.

Furthermore, the $T_H$ curve roughly extends to the peak of the superconducting dome, indicating a Fermi surface reconstruction near the maximum of $T_c$. From the observed superconducting dome, $R_H$ of samples in the higher doped regime tends to be negative at low temperature and evolves toward less negative values with increasing temperature throughout the measured range; $T_c$ of samples with positive $R_H$ at low temperature tends to decrease with increasing $R_H$ (50 K), congruent with the lower doped behavior. When $R_H$ (50 K) further increases, the dome is terminated and the superconductivity vanishes, corresponding to $R_H$ (50 K) ‰ $0.155 \times 10^{-3}$ $cm^3$ $C^{-1}$ in our experiments. The datapoints of $T_c$ vs $R_H$ (40 K) in $La_{2.85}Pr_{0.15}Ni_2O_7$ thin film and $La_{2.5}Pr_{0.5}Ni_2O_7/La_{2.85}Sr_{0.15}Ni_2O_7$ heterostructure follow great consistency with the dome [21, 69]. It should be noted that the thinner $La_3Ni_2O_7$ and $La_2PrNi_2O_7$ films featuring an additional $(La,Pr)_2NiO_4$ layer at the interface generally exhibit larger $R_H$ values [20, 22, 70], suggesting the extra conducting channel may contribute to the Hall measurements in such ultrathin films [71].

Notably, it is unlikely that lattice degradation contributes to the suppression of superconductivity, since XRD $2\theta$-$\omega$ scans in Fig. S1 and rocking curves in Fig. S6 confirm the preserved sample quality [55]. The estimated phase purity through Scherrer fits remains nearly 100% during the whole annealing processes [Fig. S2(h)], further indicating the maintained crystalline quality [55, 72]. Superconductivity could also be reversibly restored through another ozone annealing process, as demonstrated in Fig. S2 [55]. On the other hand, the $c$-axis lattice constant varies within only 0.5%, without notable relationship observed between $T_{c,98\%}$ and $c$-axis lattice constant across all the studied Sr doping range as depicted in Fig. S7, which is consistent with previous report [20, 55, 68].

Intriguingly, the sign change of $R_H$ (50 K)—from negative to positive upon oxygen depletion—reveals an electron-hole crossover, which contradicts the intuitive expectation that hole doping would shift Hall effect from electron-like to hole-like behavior. We demonstrate that this behavior highly mirrors that observed in electron-doped cuprates, in which the $R_H$ changes sign from negative to positive at low temperature with increasing Ce doping or decreasing oxygen content [48, 73-75]. It is believed that this counterintuitive sign change is attributed to a Fermi surface reconstruction near a quantum phase transition, where density wave ordering breaks the Fermi surface into electron and hole pockets near optimal doping, with these two pockets connecting to form a large hole-like Fermi surface upon further electron doping [73, 74, 76-80]. Such mechanism could also govern the electronic behavior in bilayer nickelates, which warrants further investigations of the Fermi surface and the phase diagram. We also note that in extremely overdoped $La_{2-x}Sr_xCuO_4$ ($x$ ‰ 0.3), an anomalous zero-crossing to negative of $R_H$ with increasing hole doping is observed, coinciding with a Lifshitz transition of the Fermi surface [57, 81, 82].

Moreover, the $T_H$ curve roughly extending to the maximum of $T_c$ in the superconducting dome, as depicted in Fig. 4(a), aligns with the behavior in electron-doped cuprates, where the sign change in the $R_H$ coincides with the optimal $T_c$ in both Ce doping and oxygen control cases [48, 73, 74, 83]. Another phase diagram in $Pr_{2-x}Ce_xCuO_4$ analogously employs $R_H$ to trace the carrier density and illustrates the evolution of superconductivity, clearly demonstrating that the highest $T_c$ is consistently observed near $R_H = 0$ [84], which further supports the similarities between bilayer nickelates and electron-doped cuprates albeit with distinct electronic structures.

In summary, we have systematically investigated the electrical transport properties of bilayer $La_{3-x}Sr_xNi_2O_{7-\delta}$ thin films across various Sr doping levels ($0 \leq x \leq 0.21$) and oxygen content. With the carrier density characterized by $R_H$ (50 K), a universal superconducting dome emerges, above which the $T$-linear resistivity and $\ln 1/T$ insulating behavior are also identified. The counterintuitive sign change in the $R_H$, marking an electron-hole crossover near the maximum of $T_c$, shares key signatures with the electron-doped cuprates, suggesting a Fermi surface reconstruction and potential quantum phase transition in superconducting bilayer nickelates, which warrants further investigations.

*Note added.*—During the preparation of this manuscript, we became aware of another report on $La_{2.85}Pr_{0.15}Ni_2O_7$/$SrLaAlO_4$ heterostructures [85], which found a superconducting-insulator transition when tuning the oxygen content through similar vacuum annealing treatment.

*Acknowledgements*—This work was supported by the National Key R&D Program of China (Grants No. 2021YFA1400400 and No. 2022YFA1402502), the Fundamental Research Funds for the Central Universities (021314380277, 021314380269), the National Natural Science Foundation of China (No. 12434002 and No. 123B2051), the Natural Science Foundation of Jiangsu Province (Grant No. BK20233001), and the Interdisciplinary Program of Wuhan National High Magnetic Field Center, Huazhong University of Science and Technology (Grant No. WHMFC2025013).

*Data availability*—The data that support the findings of this article are openly available [86].

[*]These authors contributed equally to this work.

[†]ynie@nju.edu.cn

## Figures and captions

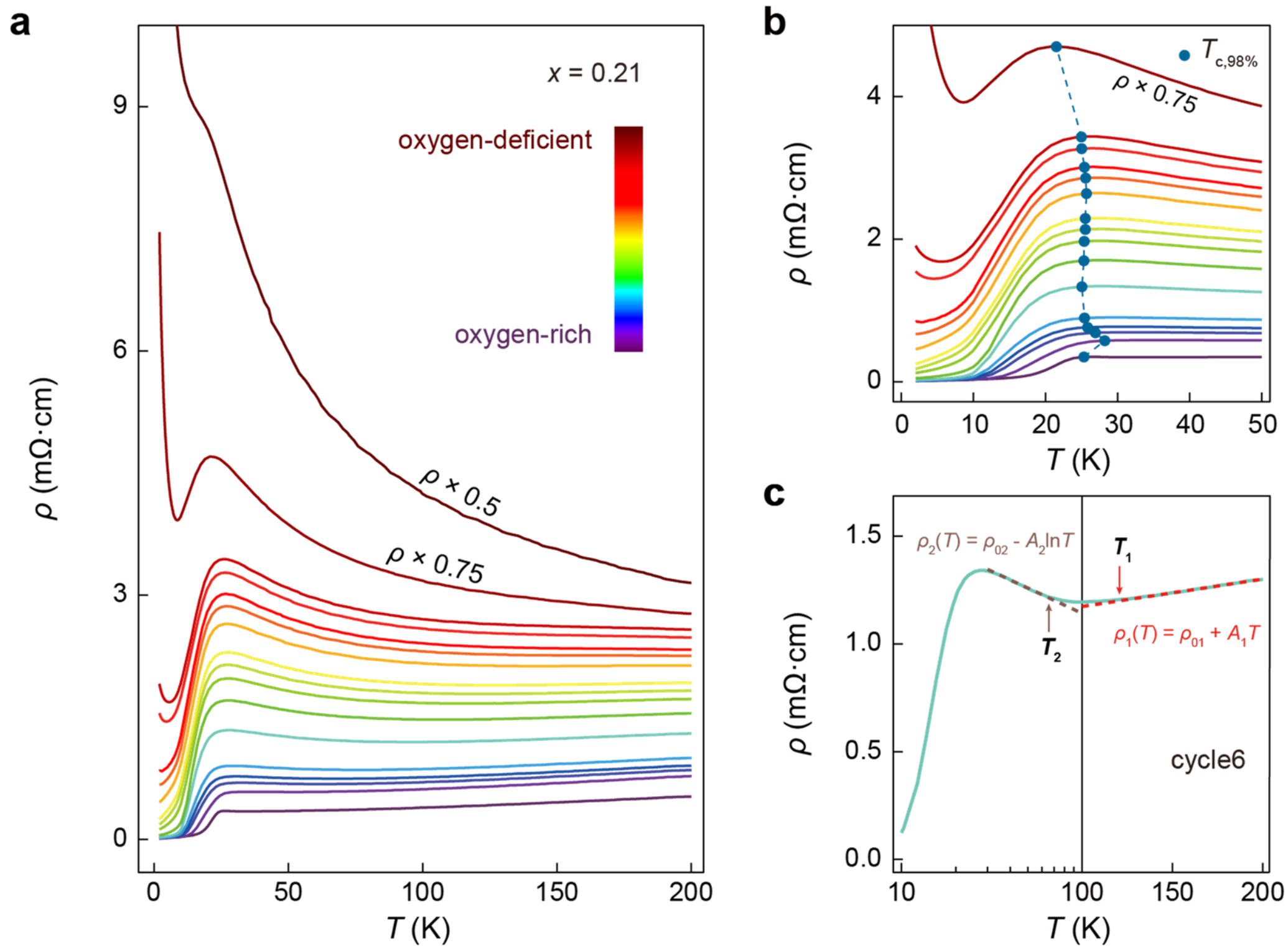

FIG. 1. Transport properties of $La_{2.79}Sr_{0.21}Ni_2O_7$ thin film with different oxygen content. (a) $\rho(T)$ curves (200-2 K) of $La_{2.79}Sr_{0.21}Ni_2O_7$ thin film. The curves from bottom to top are corresponding to the oxygen content from high to low. The oxygen content was tuned through *in situ* vacuum annealing from cycle2 to cycle17, after an initial ozone annealing for cycle1 (dark purple curve at the bottom). (b) Enlarged $\rho(T)$ curves (50-2 K) showing superconducting transition from cycle1 to cycle16 in (a). Solid blue circles represent $T_{c,98\%}$ values for each cycle, defined as the temperature at which the measured resistivity drops to 98% of the extrapolated value from a linear fit to the normal state resistivity near the onset temperature. (c) Resistivity $\rho$ for cycle6 plotted versus linear $T$ at high temperature (200-100 K) and logarithmic $T$ at low temperature (100-10 K). $T_1$ and $T_2$ are the characteristic temperatures at which the measured $\rho(T)$ deviates from the $T$-linear fit $\rho_1(T)$ (red dashed line) and $\ln 1/T$ fit $\rho_2(T)$ (brown dashed line) by 0.5%, respectively.

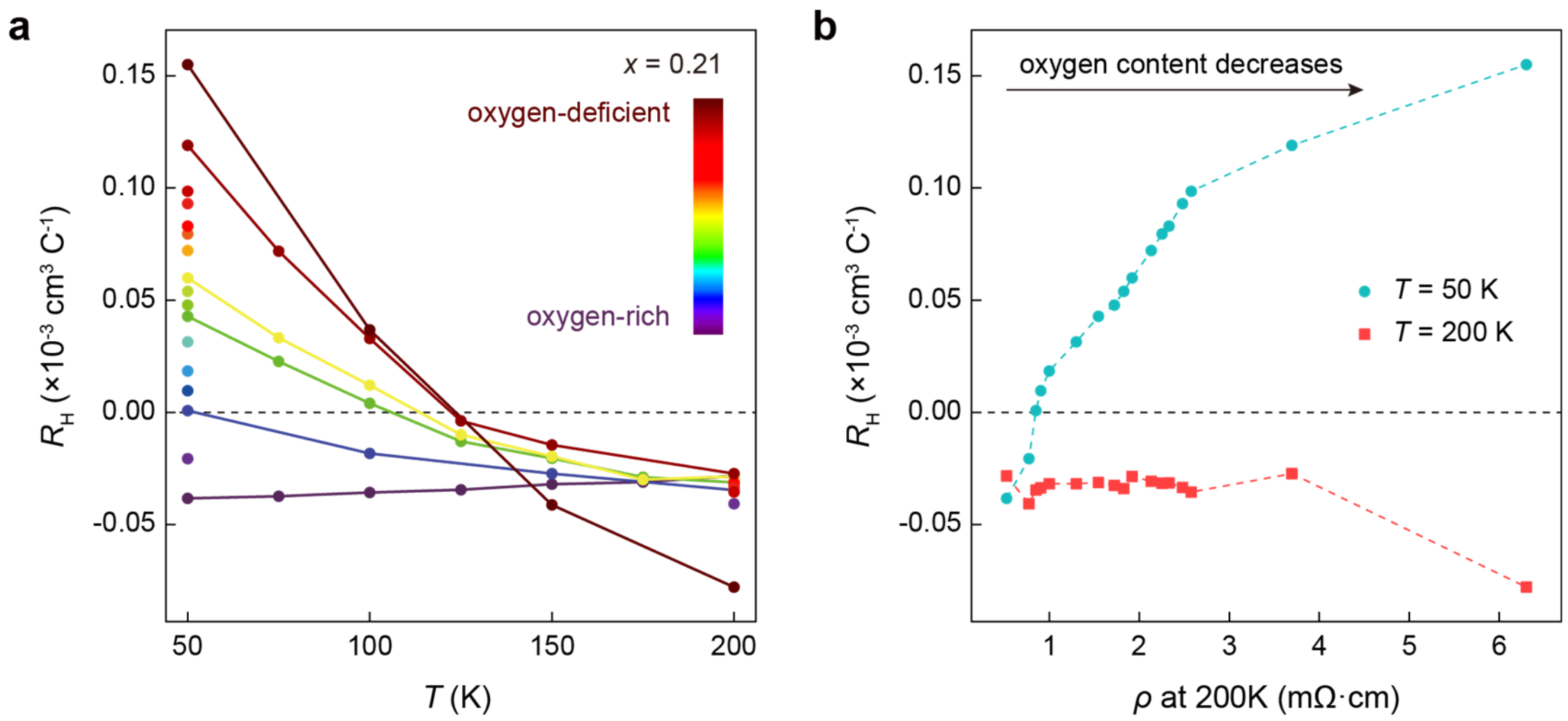

FIG. 2. Hall coefficient $R_H$ of $La_{2.79}Sr_{0.21}Ni_2O_7$ thin film with different oxygen content. (a) $R_H$ versus $T$ from 200 K to 50 K. The curves from bottom to top are corresponding to the oxygen content from high to low. (b) $R_H$ at 200 K and 50 K as a function of the resistivity $\rho$ at 200 K. The dashed lines are guides to the eye. The arrow marks the direction of oxygen content decrease.

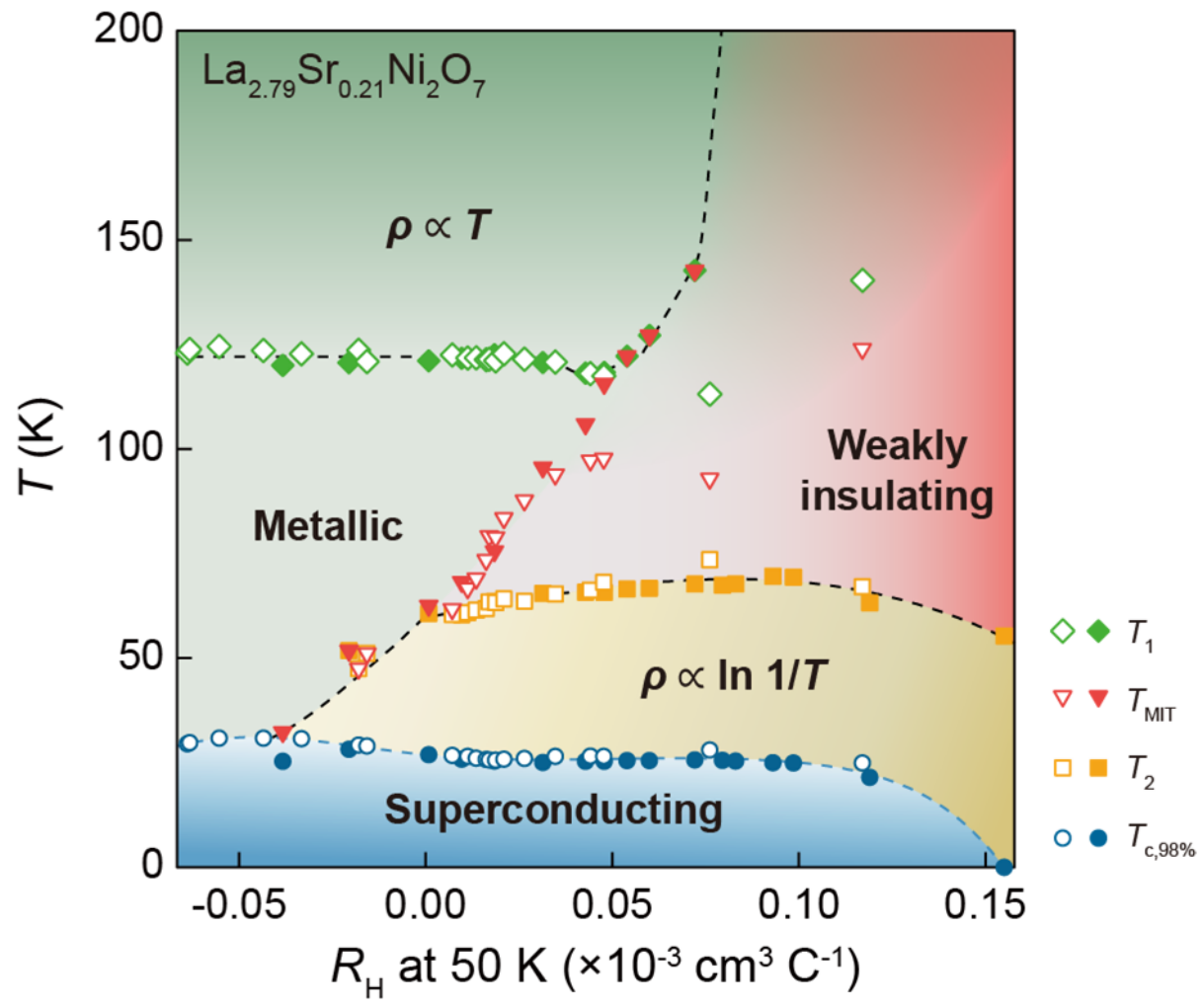

FIG. 3. $T$-$R_H$ phase diagram for electrical transport properties in $La_{2.79}Sr_{0.21}Ni_2O_7$ thin films. Green diamonds represent the temperature $T_1$ above which $\rho(T)$ follows the $T$-linear dependence; red triangles represent the metal-insulator transition temperature $T_{MIT}$; orange squares represent the temperature $T_2$ below which $\rho(T)$ exhibits an $\ln 1/T$ insulating behavior; blue circles represent the superconducting transition temperature $T_{c,98\%}$. $T_1$ and $T_2$ are defined as the temperatures at which the measured $\rho(T)$ deviates from the $T$-linear fit and $\ln 1/T$ fit by 0.5%, respectively. Hollow and solid symbols correspond to two different samples.

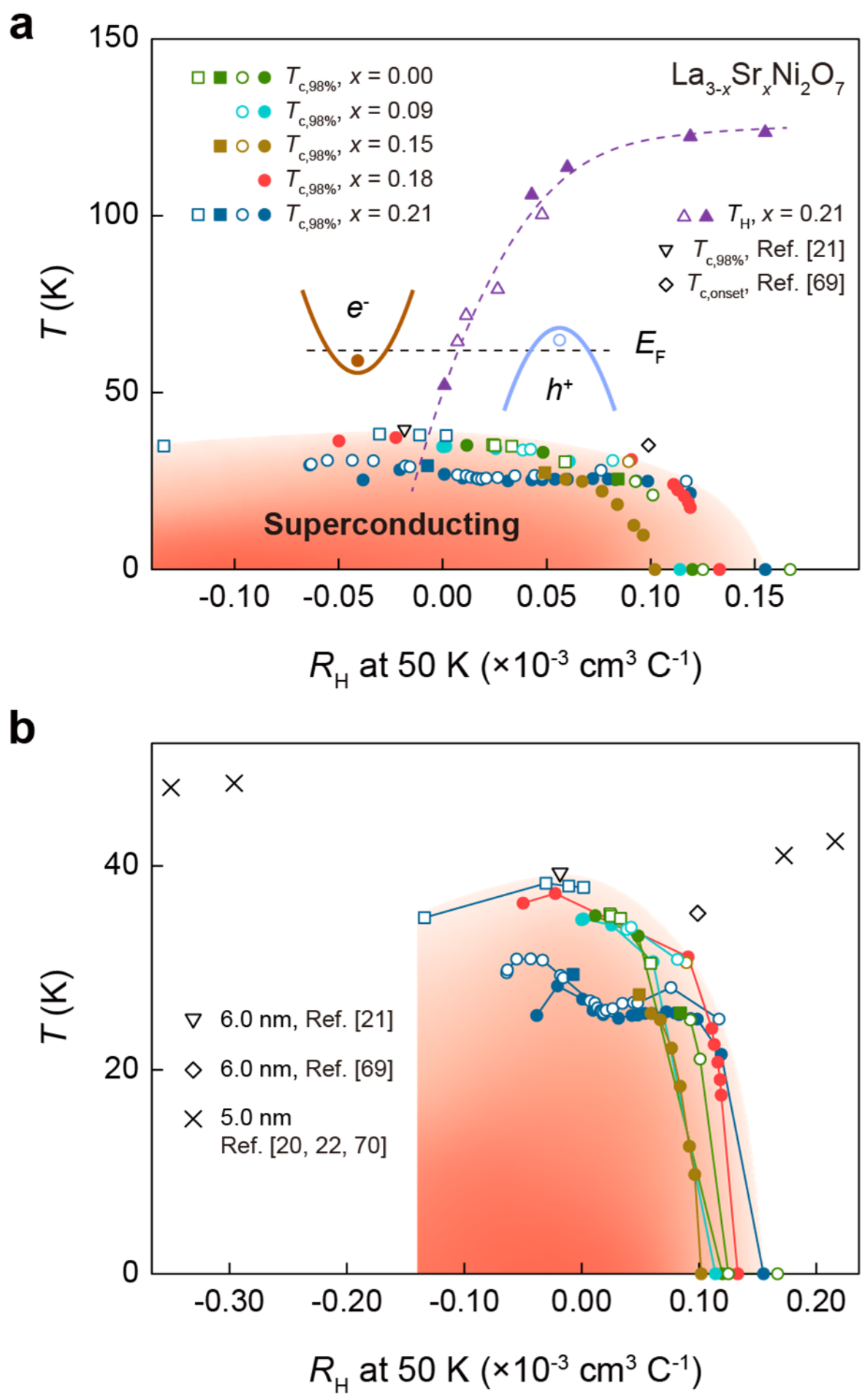

FIG. 4. Superconducting dome in $La_{3-x}Sr_xNi_2O_{7-\delta}$ thin films. (a) $T_{c,98\%}$ versus $R_H$ (50 K), with symbols representing individual samples and color indicating Sr doping level. $T_H$ denotes the electron-hole crossover temperature, defined as where $R_H(T)$ curves cross the zero axis in Fig. 2(a) and Fig. S4(m) [55]. Inset demonstrates a schematic of a two-band model showing electron ($e^-$) and hole ($h^+$) bands dominating on either side of the $T_H$ boundary (purple dashed line). The black dashed line represents the Fermi level $E_F$. (b) Comparison between the proposed superconducting dome and the datapoints from other studies [20-22, 69, 70]. Lines connect data from the same sample with different oxygen content. $R_H$ from Ref. [21, 69] is taken at 40 K. The deviation in 5.0 nm-thick films from Ref. [20, 22, 70] may be attributed to the reduced thickness and the contribution of an extra $(La,Pr)_2NiO_4$ layer at the interface.

# Supplementary Materials

## Superconducting Dome in $La_{3-x}Sr_xNi_2O_{7-\delta}$ Thin Films

Maosen Wang,[1,2,*] Bo Hao,[1,2,*] Wenjie Sun,[1,2,*] Shengjun Yan,[1,2] Shengwang Sun,[1,2] Hongyi Zhang,[1,2] Zhengbin Gu,[1,2] and Yuefeng Nie[1,2,3,†]

[1]National Laboratory of Solid State Microstructures, Jiangsu Key Laboratory of Artificial Functional Materials, College of Engineering and Applied Sciences, Nanjing University, Nanjing 210093, China.

[2]Collaborative Innovation Center of Advanced Microstructures, Nanjing University, Nanjing 210093, China.

[3]Jiangsu Physical Science Research Center, Nanjing 210093, China.

[*]These authors contributed equally to this work.

[†]ynie@nju.edu.cn

**This file includes:**

I. Experimental Methods

II. Additional experimental results, Figs. S1-S11

# I. Experimental Methods

**Epitaxial film growth.** High quality $La_{3-x}Sr_xNi_2O_{7-\delta}$ thin films with various Sr doping levels ($x$ = 0.00, 0.09, 0.15, 0.18, 0.21) were synthesized on single-crystalline (001) $SrLaAlO_4$ substrates using a DCA R450 reactive molecular-beam epitaxy (MBE) system with distilled ozone as oxidizing agent. $SrLaAlO_4$ substrates were pre-annealed at 1000°C in air for 2 hours with $LaAlO_3$ substrates covered before growth using a HF-Kejing KSL-1700X box furnace. The substrate temperature was fixed at 720°C during growth with a background pressure of ~$1 \times 10^{-5}$ Torr. A shuttered deposition method was used in the layer sequence of [(La,Sr)O] - [(La,Sr)O] - [$NiO_2$] - [$NiO_2$] - [(La,Sr)O] with the deposition process and surface quality monitored via *in situ* reflection high-energy electron diffraction (RHEED). For all samples, the thickness of nickelate films was maintained at 3 unit cells (u.c.), i.e., 6 *n*-layer blocks with the estimated thickness of 6.2 nm, before being capped with 1 u.c. $SrTiO_3$ layer serving as structural support. After growth, the films would not be transferred out of the growth chamber until cooled down to room temperature with the oxidizing background pressure unaltered to avoid inducing additional oxygen vacancies.

**X-ray diffraction scans.** The x-ray diffraction $2\theta$-$\omega$ scans were performed using a Bruker D8 Discover diffractometer with a monochromated Cu $K_{\alpha 1}$ ($\lambda$ = 1.5406 Å) radiation source to evaluate the crystalline quality.

**Ozone annealing.** The as-grown films were post-annealed at 380°C for 1 hour in a reactive ozone atmosphere achieved via a commercial ozone generator (AC-2025, IN USA Inc.). The oxygen flow rate into the ozone generator was maintained at ~100 sccm during the whole annealing process, while the ozone generator output was fixed at 10.5% in the temperature holding period, deliberately exceeding the optimal parameters to achieve oxygen overdoped films. Both the warming-up and cooling-down processes were performed as fast as possible to suppress the possible phase transition towards higher-order RP phases or oxygen-intercalated layered perovskites [1-3], with the time length limited to 3 minutes and 6 minutes, respectively.

**Electrical transport measurements.** DC electrical transport properties were measured using a Physical Property Measurement System (PPMS, Quantum Design) in Hall bar configurations, with a current of 5 μA for resistivity measurements and 500 μA for Hall effect measurements. The Hall bar configurations were deposited with Au electrodes via DC sputtering and patterned with hard shadow masks, after which a diamond scribe was employed to cut the sample into ~5 mm × 5 mm pieces and the ohmic contacts were made by ultrasonic aluminum wire bonding. The oxygen mis-stoichiometries introduced in the sputtering process could be eliminated through post-ozone annealing. To ensure the oxygen content at each measurement was steady, the temperature-dependent resistivity was measured below 200 K [4].

**Oxygen content modulation.** After the initial excess ozone annealing, a series of in situ vacuum annealing processes were implemented to modulate the oxygen content systematically in the PPMS chamber with a background of ~10 Torr helium gas. The annealing temperature and time gradually increase for each cycle as vacuum annealing progresses. For the representative sample in Fig. 1, the annealing temperature increases from 370 K for the first few cycles to 400 K for the last few cycles, while the annealing time increases from around 10 min to 180 min. The current was also applied above 300 K during the vacuum annealing process to achieve the real-time monitoring of the resistivity variation.

# II. Additional experimental results

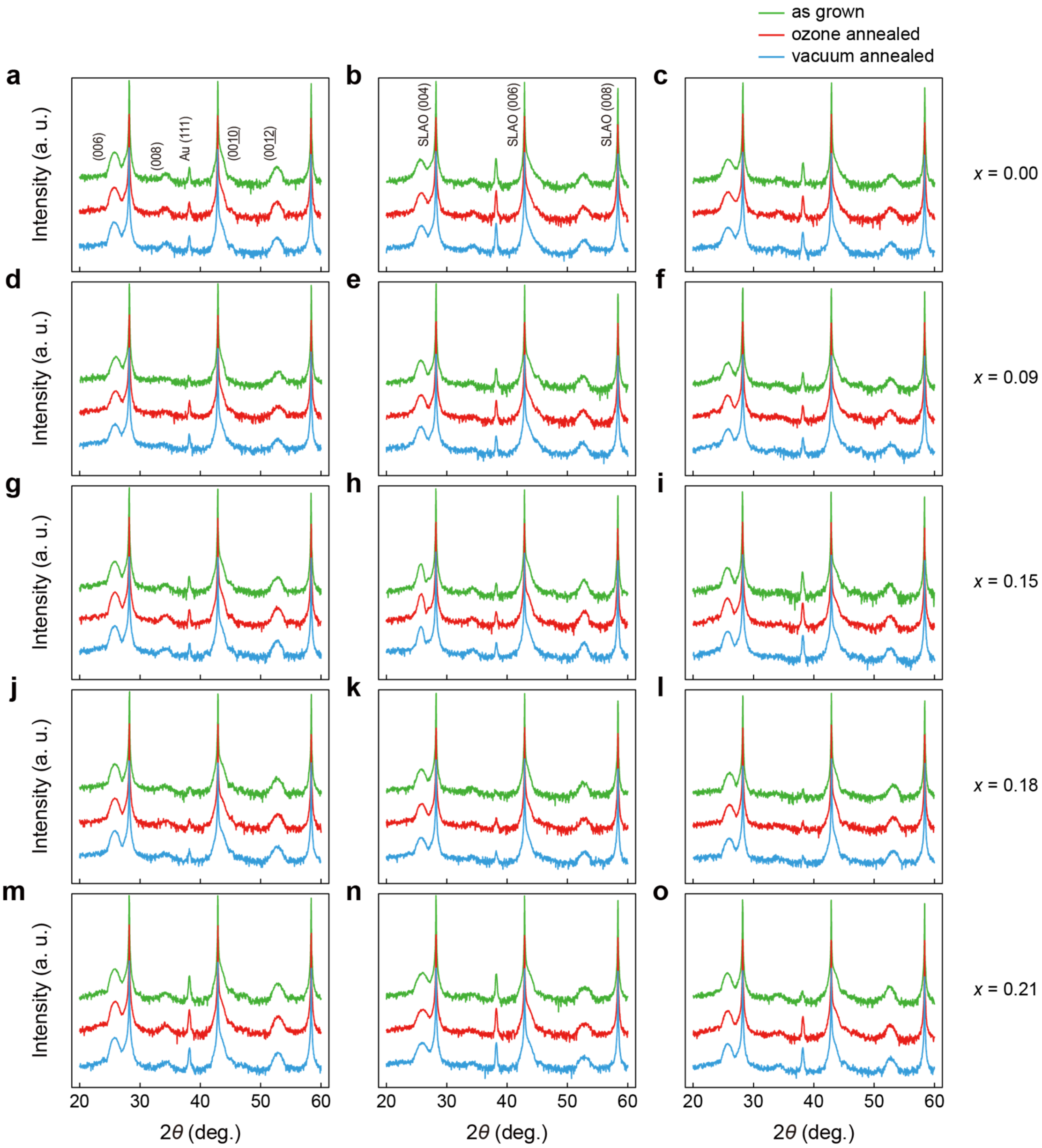

FIG. S1. X-ray diffraction $2\theta$-$\omega$ scans of $La_{3-x}Sr_xNi_2O_{7-\delta}$ thin films with different oxidation states grown on $SrLaAlO_4$ (001) substrate. Three different samples are shown for each Sr doping level. Green, red and blue profiles represent the as-grown, ozone annealed and vacuum annealed state, respectively. The post-growth ozone annealing and subsequent vacuum annealing process produce negligible influence on the crystalline quality.

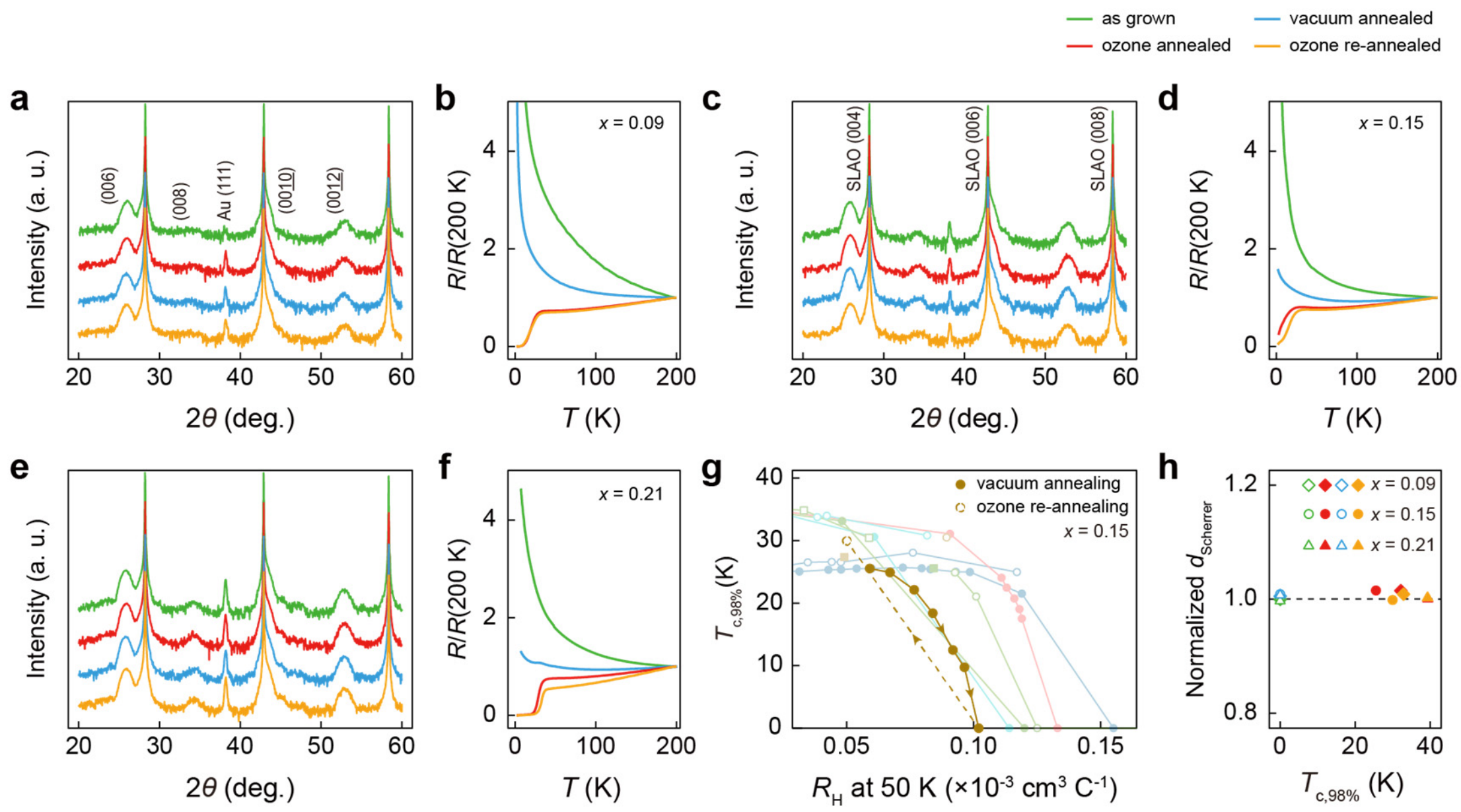

FIG. S2. Reversibility of the vacuum annealing process. (a), (c), (e) XRD $2\theta$-$\omega$ scans of $La_{3-x}Sr_xNi_2O_{7-\delta}$ thin films ($x$ = 0.09, 0.15, 0.21) with different oxidation states. (b), (d), (f) Corresponding $\rho(T)$ curves (200-2 K) of different Sr doping levels. (g) The superconducting transition temperature $T_{c,98\%}$ versus $R_H$ at 50 K for the sample with $x$ = 0.15 in (c) and (d). Solid and open circles denote the vacuum annealing and ozone re-annealing processes, respectively. The faded points correspond to the same points in Fig. 4. (h) The phase purity versus $T_{c,98\%}$ for the samples in (a)-(f). The phase purity is estimated by the normalized Scherrer thickness $d_{\mathrm{Scherrer}}/d_{\mathrm{Scherrer,\ as\text{-}grown}}$. Different shapes of symbols denote different Sr doping levels, while different colors represent different oxidation states.

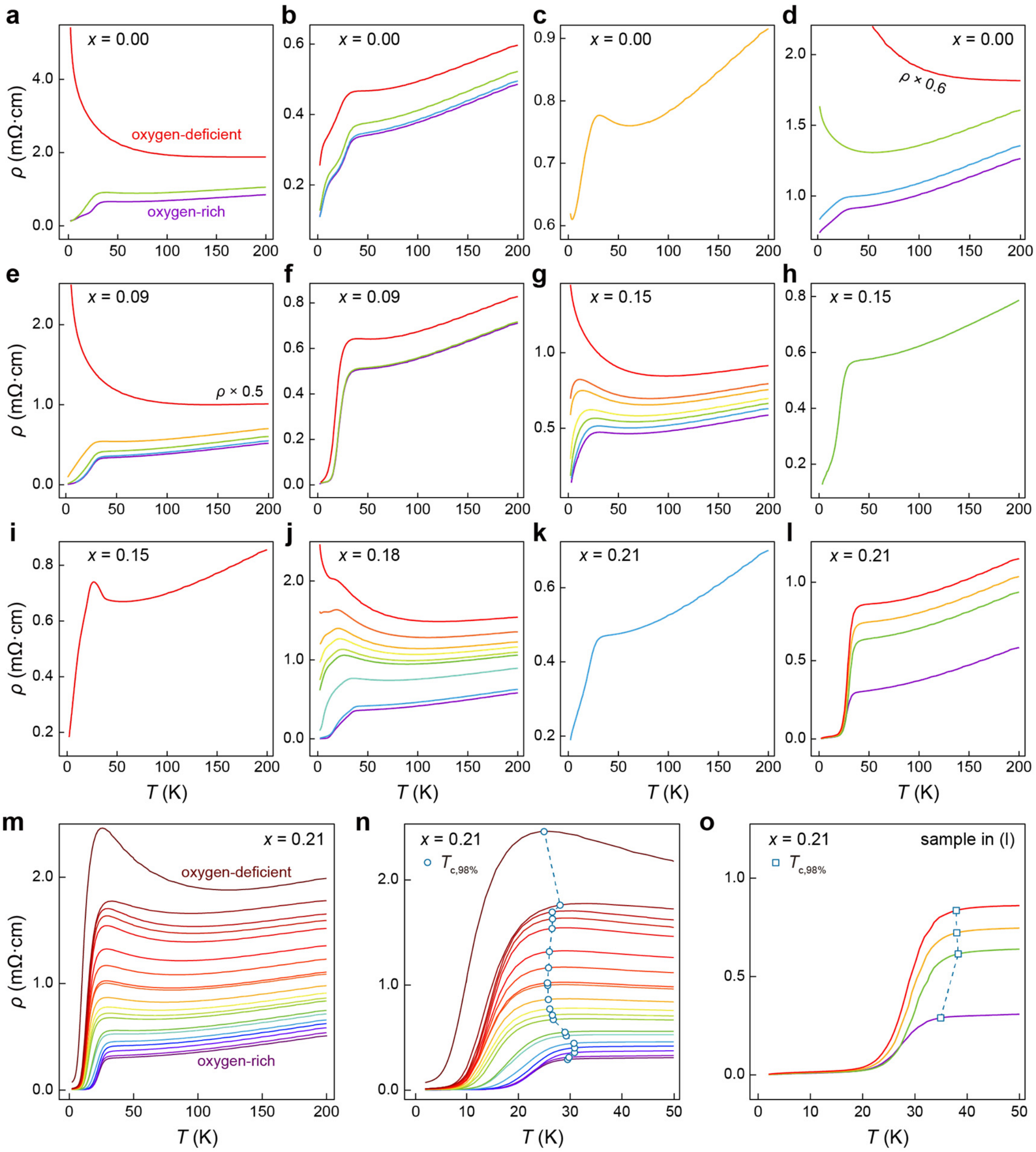

FIG. S3. Transport properties of other $La_{3-x}Sr_xNi_2O_{7-\delta}$ thin films in Fig. 4. (a)-(m) $\rho(T)$ curves measured from 200K to 2K. The curves in the same panel represent the same sample, from bottom to top corresponding to the oxygen content from high to low. (n), (o) Enlarged $\rho(T)$ curves (50-2 K) showing superconducting transition for the samples in (m) and (l), respectively. Open blue circles and squares denote $T_{c,98\%}$ values for each vacuum annealing cycle.

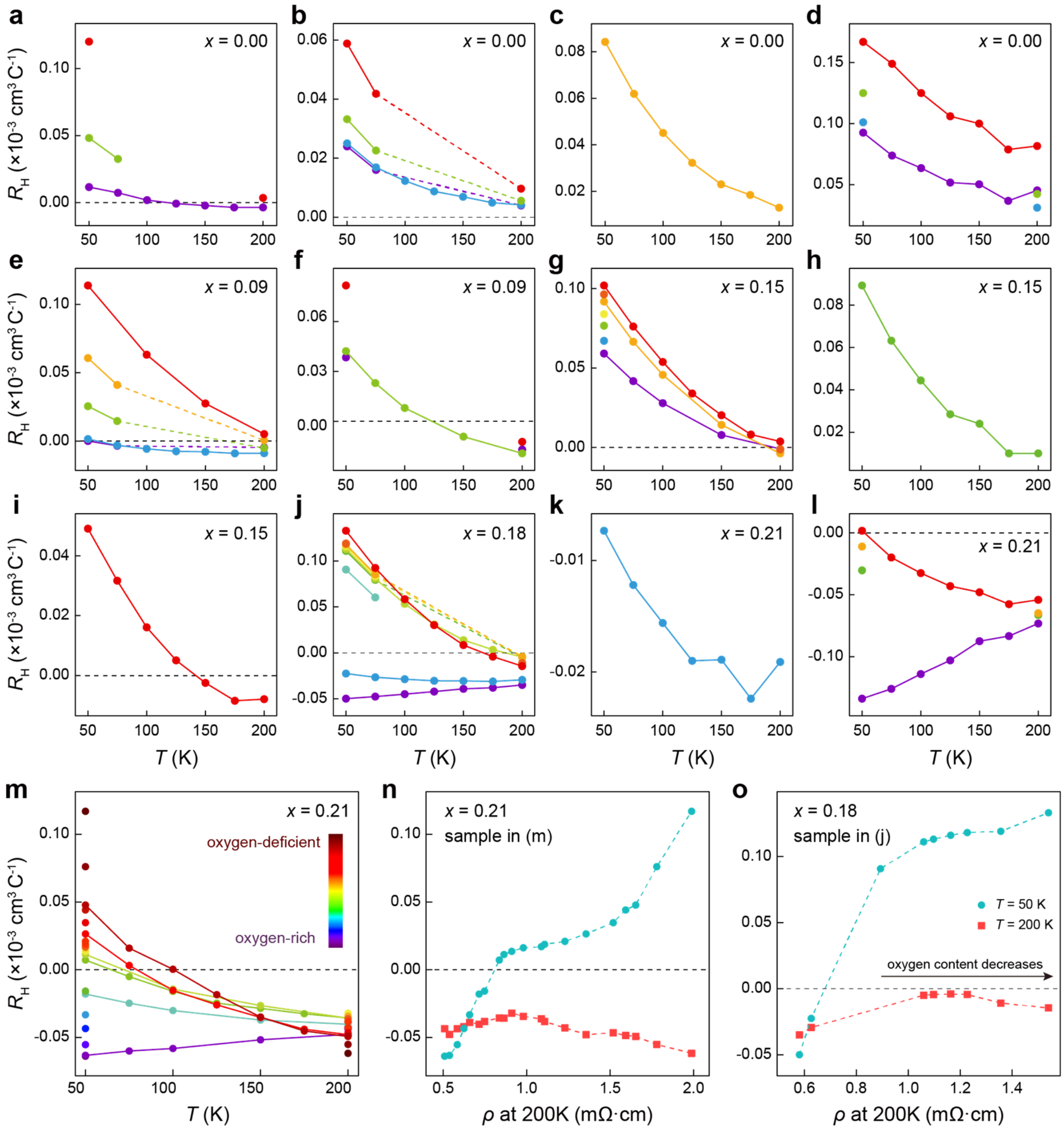

FIG. S4. Hall coefficient $R_H$ of other $La_{3-x}Sr_xNi_2O_{7-\delta}$ thin films in Fig. 4. (a)-(m) $R_H$ versus $T$ from 200 K to 50 K. The curves in the same panel represent the same sample, with the $R_H$ (50 K) points from bottom to top corresponding to the oxygen content from high to low. (n), (o) $R_H$ at 200 K and 50 K as a function of the resistivity $\rho$ at 200 K for the samples in (m) and (j), respectively. The dashed lines are guides to the eye. The arrow marks the direction of oxygen content decrease.

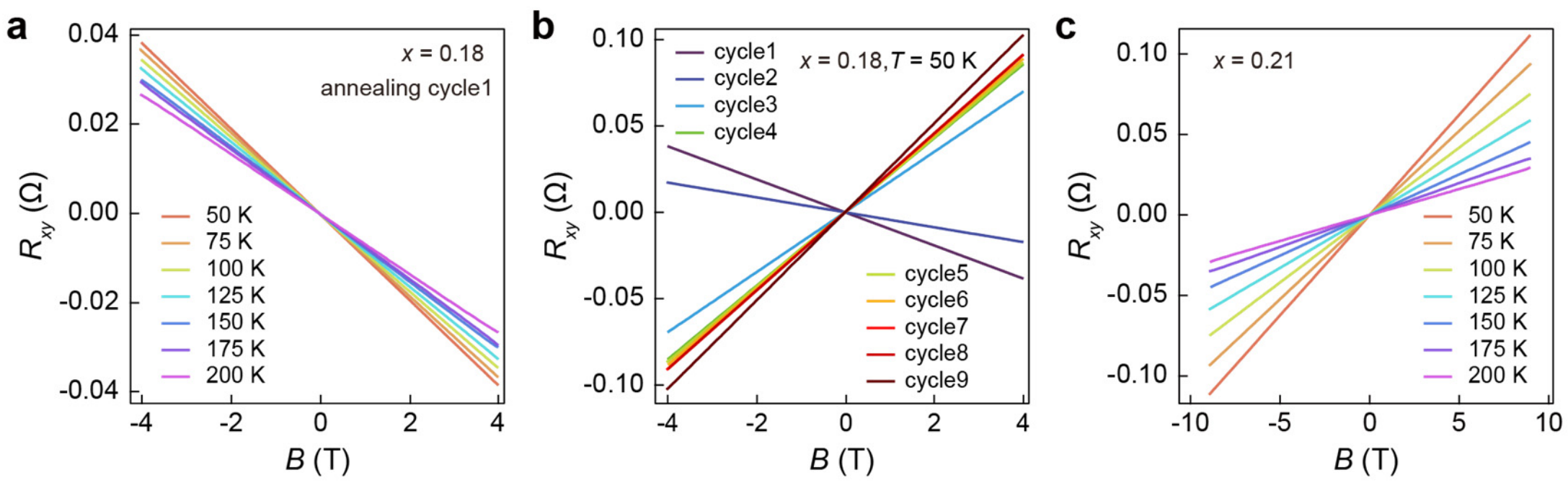

FIG. S5. Hall resistivity $R_{xy}$ versus magnetic field $B$ of a $La_{2.82}Sr_{0.18}Ni_2O_7$ thin film measured at (a) different temperatures for cycle1 before the vacuum annealing process and (b) 50 K for different oxygen content. (c) $R_{xy}$ versus $B$ up to 9 T of a $La_{2.79}Sr_{0.21}Ni_2O_7$ thin film for different temperatures. All the curves show great linear behavior.

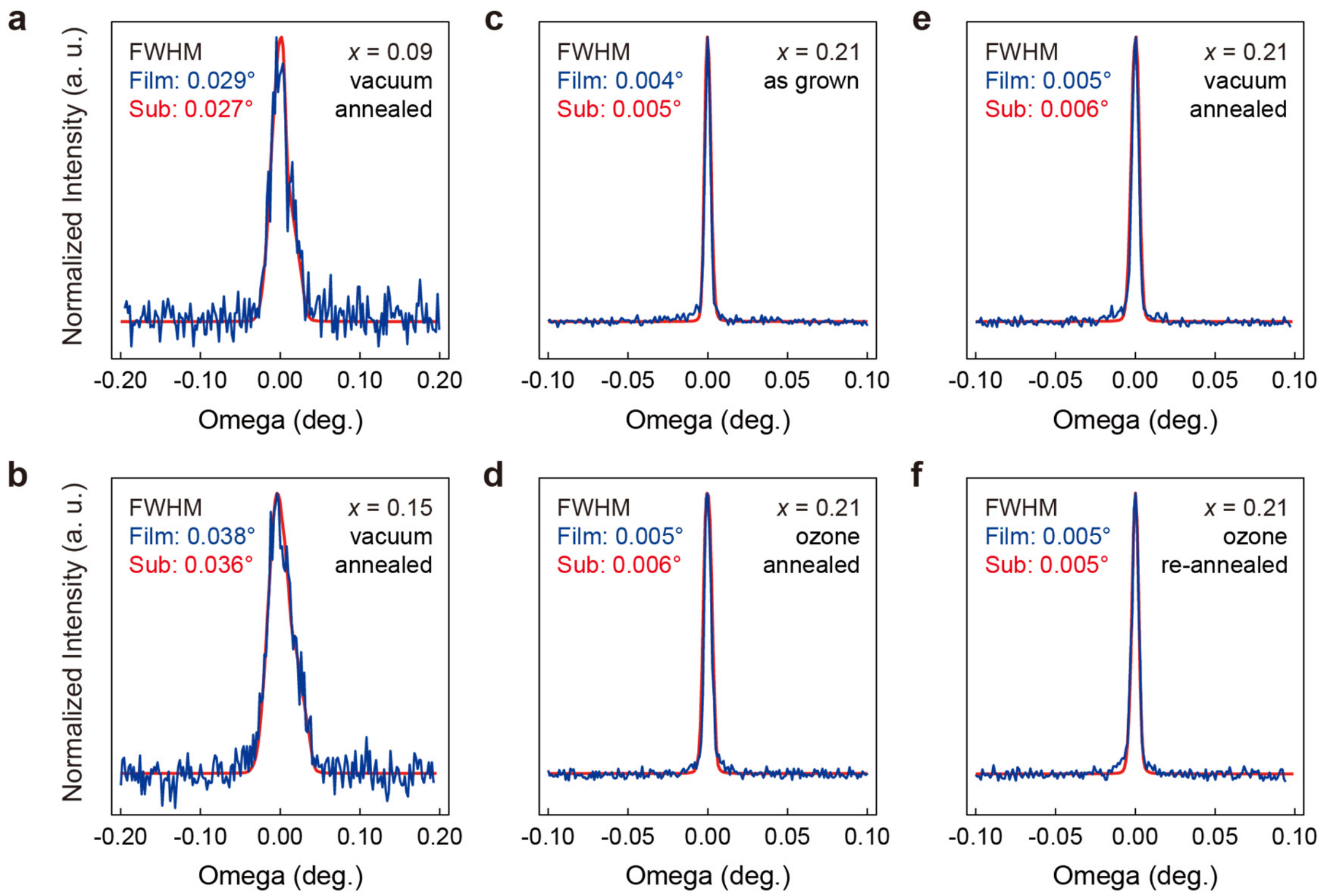

FIG. S6. Rocking curves of the $SrLaAlO_4$ (004) and $La_{3-x}Sr_xNi_2O_{7-\delta}$ (006) reflections with (a) $x$ = 0.09, (b) $x$ = 0.15 under vacuum annealed state and (c)-(f) $x$ = 0.21 with different oxidation states. The films with $x$ = 0.09, 0.15, 0.21 correspond to the samples in Fig. S3(e), (g) and (l), respectively. The rocking curves are normalized by their respective maximum. All the curves exhibit comparable full-width-at-half-maxima (FWHM) between films and substrates.

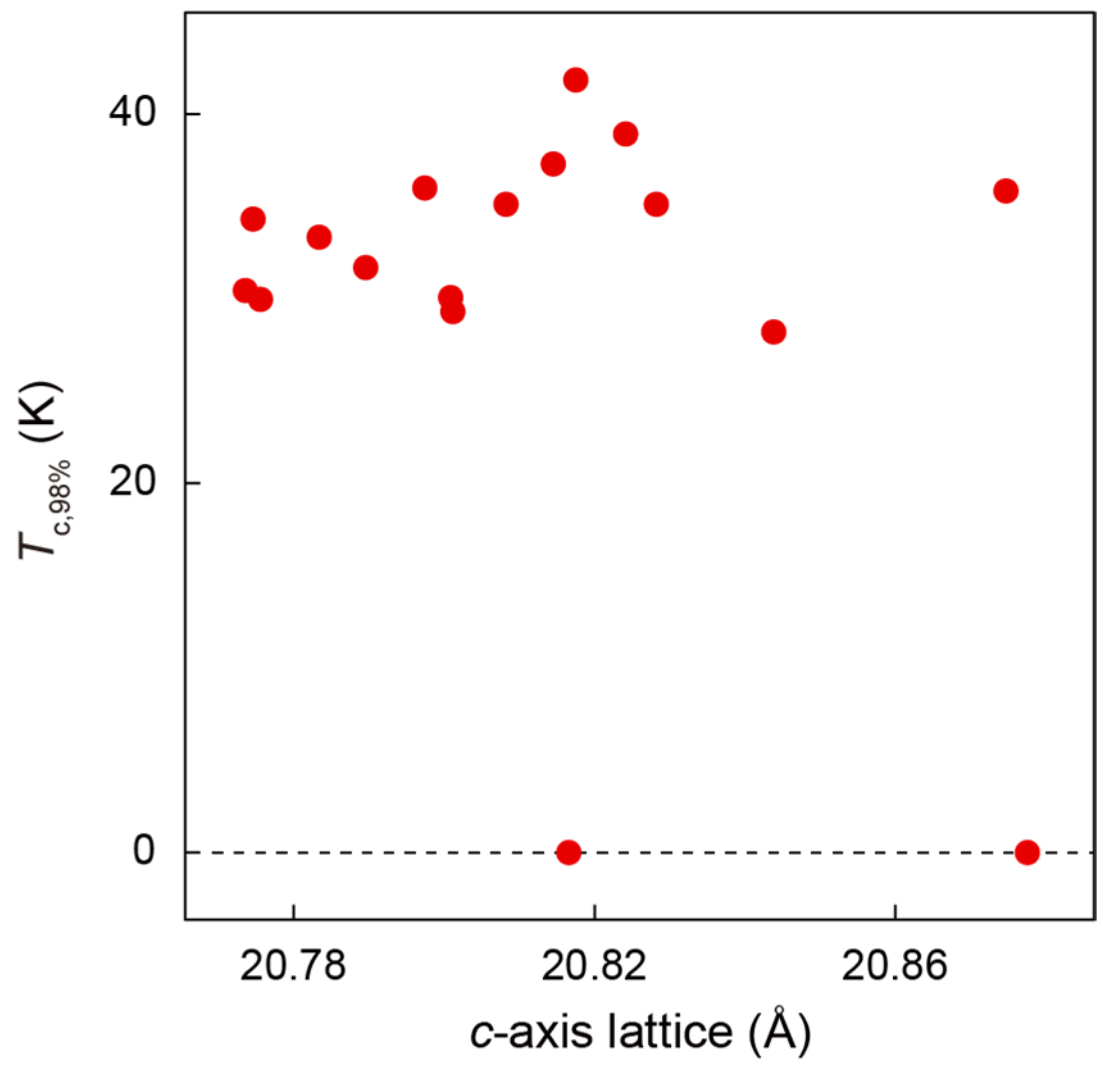

FIG. S7. $T_{c,98\%}$ versus $c$-axis lattice constant in $La_{3-x}Sr_xNi_2O_7$ thin films across all the studied Sr doping range.

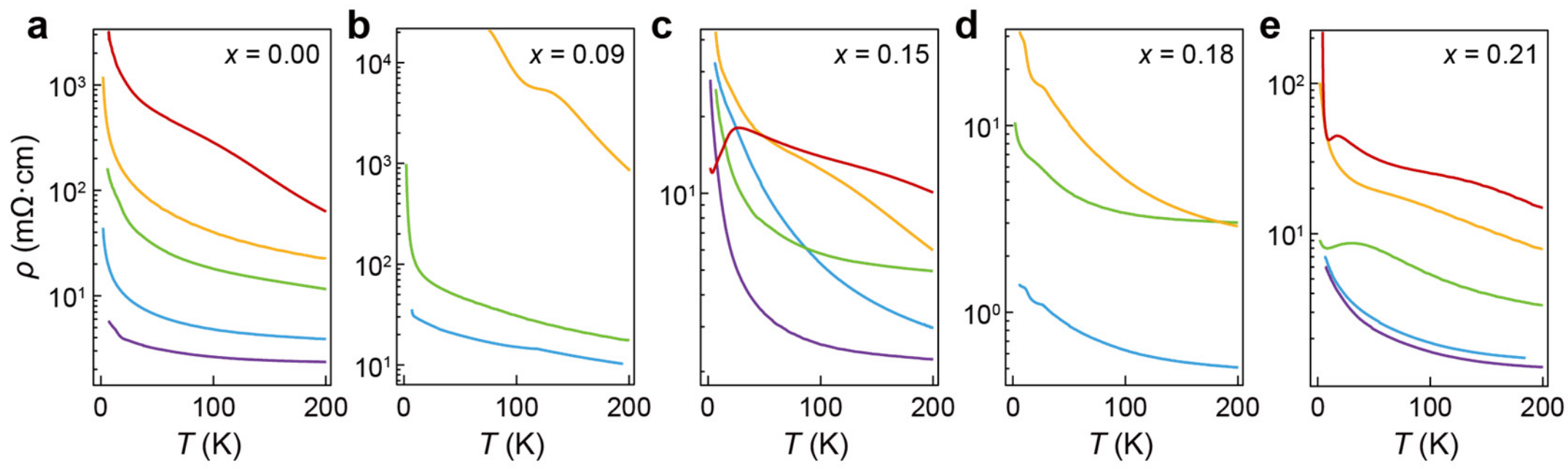

FIG. S8. $\rho(T)$ curves of as-grown $La_{3-x}Sr_xNi_2O_{7-\delta}$ thin films with (a) $x = 0.00$, (b) $x = 0.09$, (c) $x = 0.15$, (d) $x = 0.18$ and (e) $x = 0.21$. The different colors denote different samples.

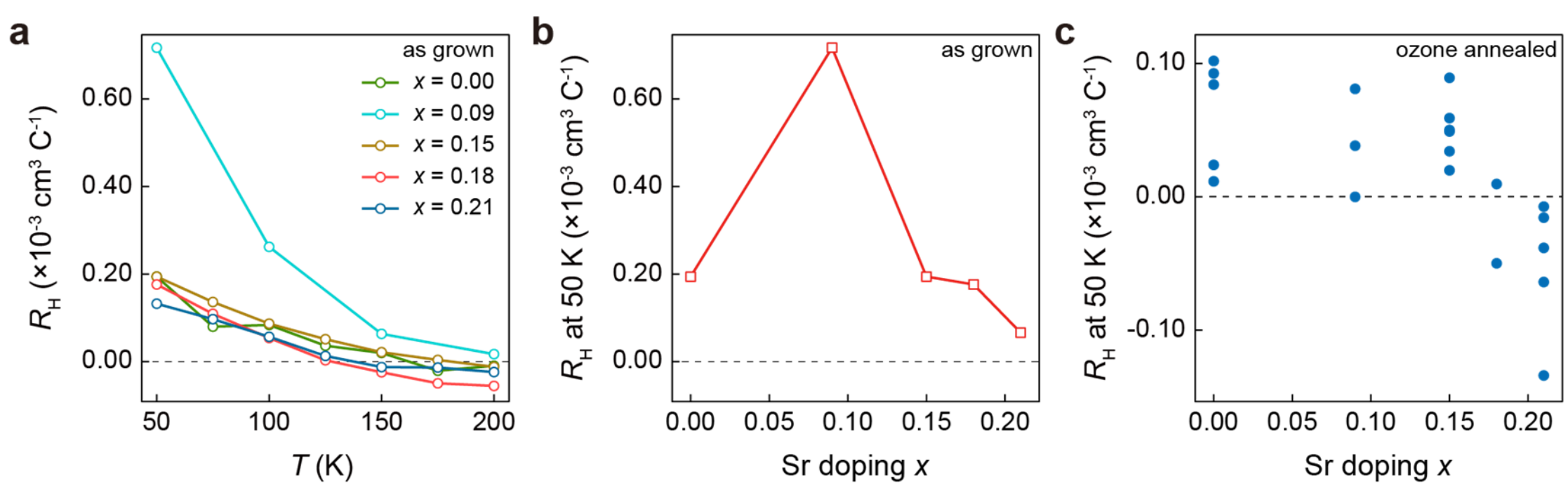

FIG. S9. Hall coefficient $R_H$ of $La_{3-x}Sr_xNi_2O_{7-\delta}$ thin films. (a) $R_H$ versus $T$ from 200 K to 50 K with different Sr doping levels for as-grown films. (b) $R_H$ at 50 K as a function of Sr doping levels $x$ for the as-grown films in (a). (c) Relation between $R_H$ at 50 K and Sr doping levels $x$ for ozone annealed films.

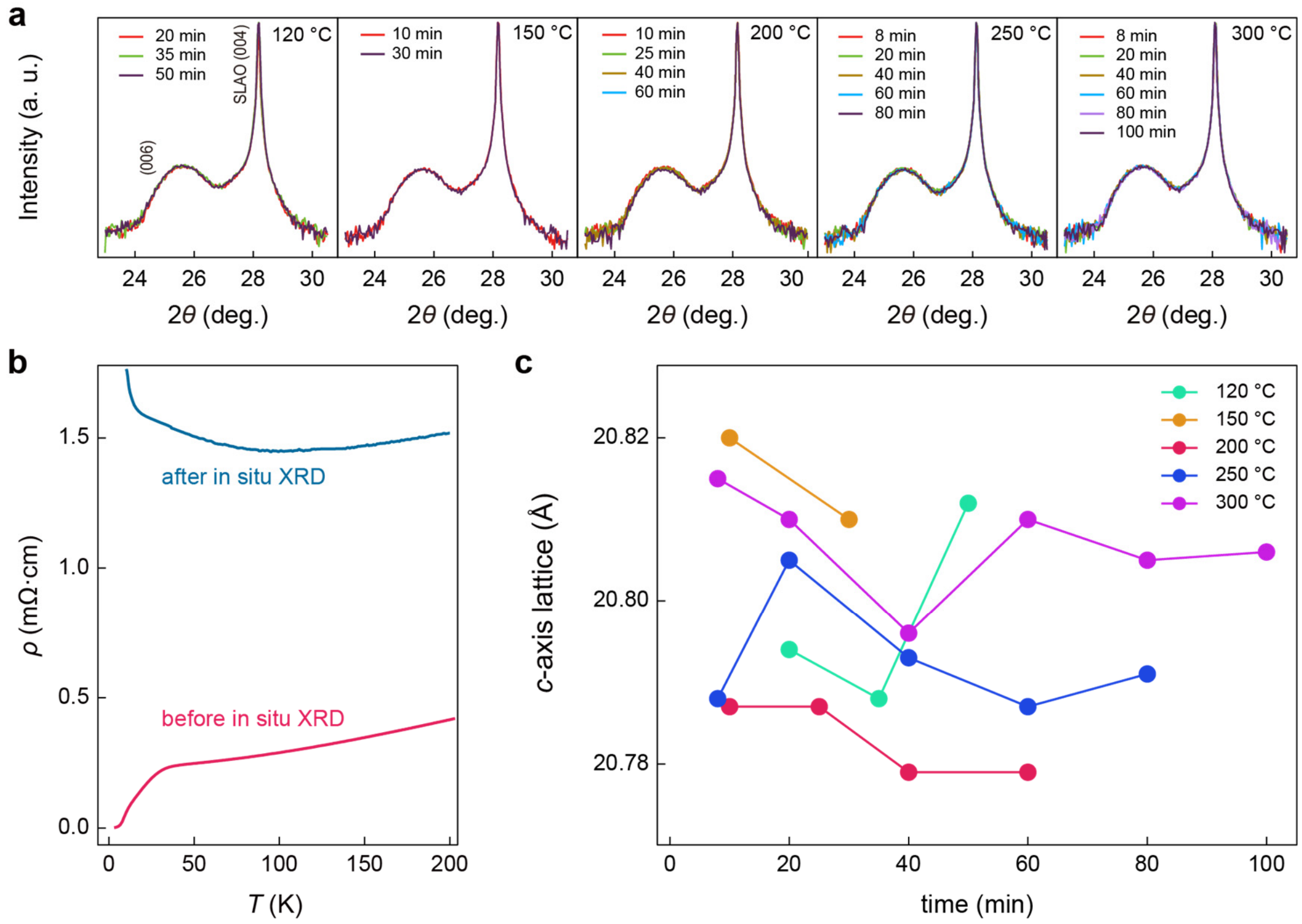

FIG. S10. In situ high temperature XRD measurements at atmosphere. (a) (006) diffraction peaks of a $La_{2.91}Sr_{0.09}Ni_2O_{7-\delta}$ thin film at different temperature and testing time. (b) $\rho(T)$ curves measured from 200K to 2K before and after the in situ XRD measurements. The film has already lost superconductivity with oxygen loss. (c) *c*-axis lattice constants versus testing time at different temperature, which exhibits no notable evolution.

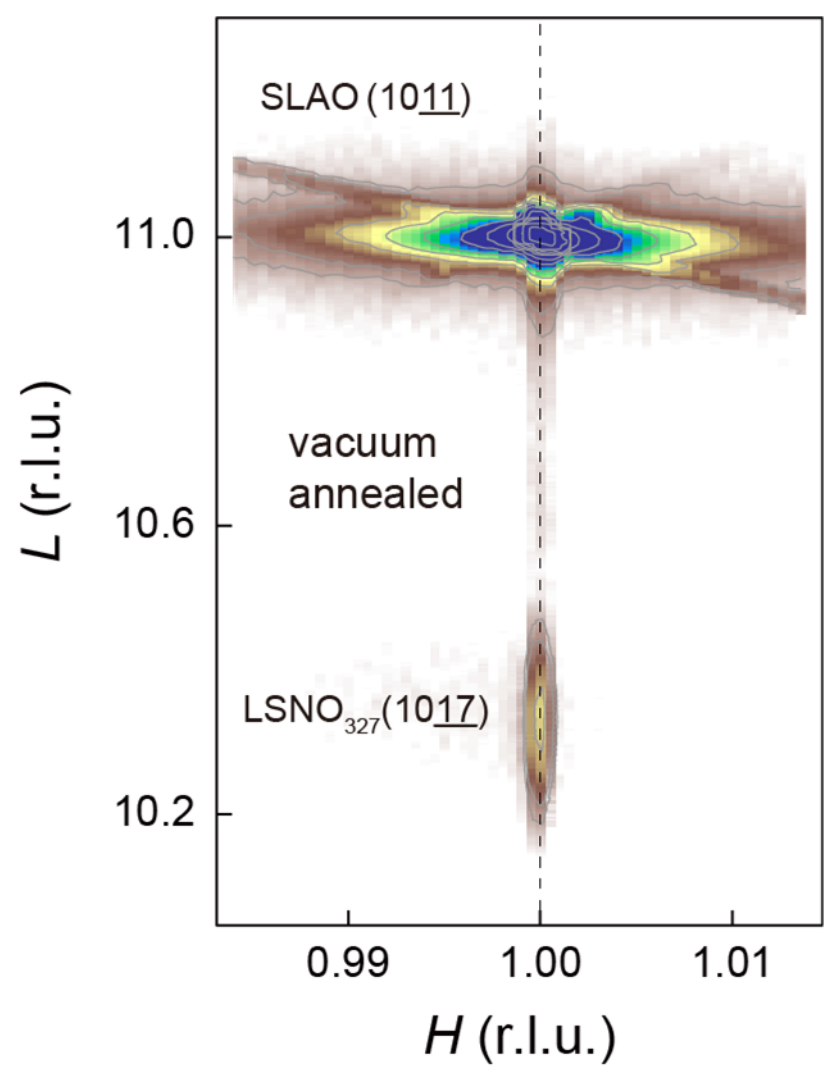

FIG. S11. Reciprocal space mapping around the $SrLaAlO_4$ (10<u>11</u>) and $La_{2.79}Sr_{0.21}Ni_2O_7$ (10<u>17</u>) reflections after vacuum annealing. The film is still coherently strained to the substrate.